\documentclass[fleqn,usenatbib,useAMS]{mnras}

\usepackage[T1]{fontenc}
\usepackage{graphicx}	
\usepackage{amsmath}	
\usepackage{amssymb}	
\usepackage{xspace}
\usepackage[capitalize]{cleveref}
\usepackage{hyperref}
\usepackage[separate-uncertainty=true]{siunitx}

\newcommand{\gravityspy}{\texttt{GravitySpy}\xspace}
\newcommand{\bayeswave}{\texttt{BayesWave}\xspace}
\newcommand{\celerite}{\texttt{celerite}\xspace}
\newcommand{\imrt}{\texttt{IMRPhenomT}\xspace}
\newcommand{\imrtp}{\texttt{IMRPhenomTP}\xspace}

\AtBeginDocument{\renewcommand{\vec}[1]{\mathbf{#1}}}
\newcommand{\chirpmass}{\ensuremath{\mathcal{M}^{\mathrm{d}}}}
\newcommand{\massratio}{\ensuremath{q}}
\newcommand{\chieff}{\ensuremath{\chi_{\rm{eff}}}}
\newcommand{\luminositydistance}{\ensuremath{d_{L}}}
\newcommand{\data}{\ensuremath{\vec{d}}\xspace}
\newcommand{\model}{\ensuremath{M}\xspace}
\newcommand{\parameters}{\ensuremath{\theta}\xspace}
\newcommand{\likelihood}{\ensuremath{\mathcal{L}}\xspace}
\newcommand{\prior}{\ensuremath{\pi}\xspace}

\newcommand{\qty}{\SI}

\title[Gaussian Processes for GW Astronomy]{Gaussian Processes for Glitch-robust Gravitational-wave Astronomy}

\author[G. Ashton]{
Gregory Ashton,$^{1}$\thanks{E-mail: gregory.ashton@rhul.ac.uk}
\\
Department of Physics, Royal Holloway, University of London, TW20 0EX, United Kingdom
}

\pubyear{2022}

\usepackage{newtxtext,newtxmath}
\begin{document}
\label{firstpage}
\pagerange{\pageref{firstpage}--\pageref{lastpage}}
\maketitle

\begin{abstract}
Interferometric gravitational-wave observatories have opened a new era in astronomy. The rich data produced by an international network enables detailed analysis of the curved space-time around black holes. With nearly one hundred signals observed so far and thousands expected in the next decade, their population properties enable insights into stellar evolution and the expansion of our Universe. However, the detectors are afflicted by transient noise artefacts known as ``glitches'' which contaminate the signals and bias inferences. Of the 90 signals detected to date, 18 were contaminated by glitches. This feasibility study explores a new approach to transient gravitational-wave data analysis using Gaussian processes, which model the underlying physics of the glitch-generating mechanism rather than the explicit realisation of the glitch itself. We demonstrate that if the Gaussian process kernel function can adequately model the glitch morphology, we can recover the parameters of simulated signals. Moreover, we find that the Gaussian processes kernels used in this work are well-suited to modelling long-duration glitches which are most challenging for existing glitch-mitigation approaches. Finally, we show how the time-domain nature of our approach enables a new class of time-domain tests of General Relativity, performing a re-analysis of the inspiral-merger-ringdown test on the first observed binary black hole merger. Our investigation demonstrates the feasibility of the Gaussian processes as an alternative to the traditional framework but does not yet establish them as a replacement. Therefore, we conclude with an outlook on the steps needed to realise the full potential of the Gaussian process approach.
\end{abstract}


\section{Introduction}

The emerging field of gravitational-wave astronomy is built on the back of a multi-decade effort to design and construct kilometre-scale interferometers that can measure their relative arm lengths to better than 1 part in $10^{21}$.
An international network of such detectors (Advanced LIGO, Virgo, and KAGRA \citep{LIGO,Virgo,KAGRA}) are now in operation and have so far observed 90 compact binary coalescences (CBC) gravitational-wave signals, including binary black hole, neutron star-black hole, and binary neutron star mergers \citep{GWTC3}.
However, of these detections, 18 are contaminated by transient non-Gaussian detector artefacts known as \emph{glitches}.
If left unaddressed, these contaminating noise sources bias astrophysical inferences, undermining the scientific outputs from this exquisite date \cite{Powell:2018csz, Macas:2022afm}.
In this work, we develop a novel solution to analyse signals contaminated by glitches, simultaneously modelling the glitch using a Gaussian process \citep{Rasmussen2004}.

Glitches represent particular epochs in which a single detector misbehaves, often due to local environmental disturbances.
Therefore, it is generally assumed that any glitch-generating process is independent between distinct detectors.
This independence is essential for providing a means to distinguish them from astrophysical signals, which produce coherent power at multiple detector sites.

The characterisation of glitches forms an integral part of analysing data from interferometric detectors for two reasons (for a review, see \citet{Davis:2022dnd}).
First, studies which can identify the source of the glitch-generating process (e.g. by temporally correlating the signal as seen in the detector strain $h(t)$ with auxiliary channels monitoring the instrument) provide a means to improve the quality of data directly resulting in a reduction in the number of glitches in future data.
Such a reduction can naturally lead to an improvement in the ability of search algorithms to identify signals. The reduced background noise rate leads to more confidence in identifying true astrophysical signals.
Second, and of relevance to this work, if a glitch overlaps a signal, care must be taken to carefully unpick the signal from the non-Gaussian noise (the glitch).

Coincidences between glitches and signals are a common problem, perhaps most distinctly demonstrated by the LIGO Livingston glitch which overlapped the first-observed binary neutron star merger GW170817 \citep{GW170817_discovery, Pankow:2018qpo}.
Two approaches were applied to the mitigation of this glitch.
A simple window function was applied for search analyses to zero out data around the glitch.
Meanwhile, for source parameter estimation, the glitch was modelled by \bayeswave \citep{Cornish:2014kda}, a flexible time-frequency wavelet reconstruction, and then a realisation of that model was subtracted from the data.
This latter method, known colloquially as ``deglitching'', has now become the de-facto standard approach to perform parameter estimation for the 20\% of signals contaminated by glitches.
While \bayeswave remains the primary tool to model the glitch, a new tool \texttt{gwsubtract} \citep{Davis:2018yrz} has also been used, which applies linear subtraction building the glitch model from a witness channel \citep{Allen:1999wh}.

\citet{Davis:2022ird} describes the basic deglitching process used to analyse glitch-contaminated signals during the third and most recent observing run of the LIGO, Virgo, and KAGRA detectors. First, a check is performed for all candidate signals to see if they admit excess glitch power. Second, the glitch is modelled, and a realisation of the glitch drawn from the model is subtracted from the data. 
Finally, the deglitched data is analysed using a standard parameter estimation framework.

Deglitching has been highly successful in dealing with contaminated signals; the most recent study \citep{Hourihane:2022doe} demonstrates its general-purpose applicability. 
However, some drawbacks motivate us to study alternative approaches.
First, the underlying algorithm that models the glitch is designed to study short-duration transient bursts.
This makes it ideally suited to deglitching short-duration ($\lesssim$\qty{1}{\s}) glitches.
However, \citet{Hourihane:2022doe} find that long-duration glitches, such as the fast and slow-scattering glitches (discussed further below), are challenging and need specialised settings.
The need for specialised settings calls for manual analyses: a slow process requiring an expert to repeatedly analyse an event and, often by eye, decide if the glitch has been adequately excised.
Second, the principle of deglitching is inherently flawed because it ignores uncertainty in the glitch model.
To deglitch the data, some draw from this model is taken. Usually, this is either the median or a random draw.
However, \citet{Hourihane:2022doe} find that there are differences in the deglitched data depending on which draw is taken for slow scattering glitches.
Such differences indicate that the glitch model has a non-negligible uncertainty which is not correctly accounted for when estimating the parameters of the astrophysical signal from the resultant deglitched data.
The solution to this problem is to analyse the signal and glitch together.
Already, \citet{Hourihane:2022doe} have demonstrated how the underlying deglitching algorithm can be extended to this end.

Proper glitch mitigation is vital for ensuring unbiased astrophysical inferences of the signal. Two clear examples of the difficulties inherent in the current process came to light during the third observing run.
The first example is the event GW200129, a binary black hole system with the largest network signal-to-noise (SNR) ratio of any observation to date \citep{GWTC3}.
Recently, it has been reported that GW200129 is the first unambiguous measurement of strong-field precession in a binary black hole system \citep{Hannam:2021pit}. 
However, the evidence is complicated by the presence of glitches removed using \texttt{gwsubtract} \citep{Davis:2022ird}.
In a recent study, \citet{Payne:2022spz} demonstrated that the evidence for spin precession depends sensitively on the glitch model: taking a different fair draw can altogether remove the evidence for spin precession.
The second example is GW191109\_010717, a heavy binary black hole with a correspondingly short-duration signal.
Time-frequency spectrograms of the data from both LIGO Hanford and Livingston show the presence of glitch artefacts overlapping the signal track.
Applying the \bayeswave algorithm, the glitches were modelled and subtracted from the data following the standard process.
However, when the deglitched data was analysed using a parameter estimation algorithm searching for evidence of beyond-General Relativity (GR) physics, the posterior distribution included multimodal features indicating a positive beyond-GR result \citep{GWTC3TGR}.
Subsequent studies demonstrated that contamination could explain this: residual noise from the glitch mimicking a beyond-GR signal.
While it has been demonstrated that \bayeswave deglitching does not contaminate the resulting data \citep{Kwok:2021zny}, this study was limited to short-duration glitches in a single detector.

The difficulties in parameter estimation and tests of GR for GW200129 and GW191109\_010717 are indicative of flaws in the current deglitching approach. 
In both cases, the presence of long-duration glitches, which are challenging to model and require manual intervention, caused issues.
Substantial effort is underway to improve the detectors ready for the next observation (O4), where it is anticipated that we will observe hundreds of binary black hole signals.
While part of that effort is also dedicated to mitigating the cause of glitches at their source, that we don't fully understand their cause suggested we will not see a significant reduction in the rate of glitches in O4.
Taking a pessimistic view that the glitch rate remains similar to previous observing runs, but the rate of signal detection increases, we will therefore expect to see an increasing number of signals contaminated (while the fraction remains approximately constant).

In this work, we introduce a new approach to gravitational-wave parameter estimation, which is fundamentally designed to improve the analysis of glitch-contaminated signals.
The core idea is to replace the traditional gravitational-wave likelihood, which assumes the data is stationary and Gaussian with a known power spectral density (PSD), with a Gaussian process (GP); see \citet{Rasmussen2004} for an introduction to Gaussian processes.
In principle, the GP approach can model arbitrary coloured noise, non-stationarity, and non-Gaussian artefacts. As we will see, this can be done using a few hyper-parameters (rather than introducing a deterministic model, which can have many thousands of parameters).

We begin in \cref{sec:methods} with an introduction to the GP approach and the limitations we apply for this feasibility study.
Then, in \cref{sec:results}, we show results applying the GP approach to three case studies. The first two demonstrate robust measurement of the parameters of a signal contaminated by a long-duration glitch.
Meanwhile, in the final case study, we discuss a novel application of the GP approach to enable time-domain tests of GR. 
Finally, in \cref{sec:conclusion}, we conclude and discuss future development needed to realise the full potential of the GP method.

\section{Methods}
\label{sec:methods}
In this section, we briefly introduce the parameter estimation problem in gravitational-wave astronomy (see also \citet{Thrane:2018qnx, Christensen:2022bxb} for recent reviews) before introducing the new GP methodology.

\subsection{Bayesian inference and gravitational-wave astronomy}
In gravitational-wave astronomy, the problem of detecting a signal is separated from measuring its properties (parameter estimation).
In this work, we concern ourselves only with the latter problem, i.e. we assume we are given some chunk of data \data (in general, this will contain data from multiple detectors) which contains an astrophysical gravitational-wave signal described by a model \model along with background detector noise. 
The standard approach taken in the field is that of Bayesian inference, namely for a set of model parameters \parameters associated with the signal model \model, we seek an approximation of the posterior distribution
\begin{equation}
    p(\parameters | \data, \model) \propto \likelihood(\data | \parameters, \model) \prior(\parameters | \model)\,,
    \label{eqn:posterior}
\end{equation}
where \likelihood and \prior are the \emph{likelihood function} and \emph{prior probability distribution} respectively and we neglect the normalizing \emph{evidence}.

The standard approach used throughout the field (see, e.g. \citet{Veitch:2014wba}) is to construct the log-likelihood function in the frequency domain.
For data from the $\ell^{\rm th}$ detector in the network consisting of a real time-series with duration $T$ and sampling frequency $f_s$, we take the discrete Fourier transform (DFT):
\begin{equation}
    \tilde{d}^{(\ell)}_j = \frac{1}{f_s}\sum_{k} d_k^{(\ell)} e^{-2\pi i j k f_s T}\,,
\end{equation}
where $j$ indexes the frequency bin and a tilde denotes a frequency-domain quantity.
Then, assuming the background noise and the signal are additive and that the noise is stationary and zero-mean with a known PSD $P(f)$, the single-detector log-likelihood is given by the \emph{Whittle likelihood}:
\begin{equation}
   \ln \likelihood(\data^{(\ell)} | \theta, \model) =
   -\frac{1}{2}\sum_{j}\ln(2\pi P_j) - \frac{2}{T} \sum_j \frac{\left|\tilde{d}_j^{(\ell)} - \tilde{\mu}_j(\parameters)\right|^2}{P_j}\,.
   \label{eqn:whittle}
\end{equation}
where $j$ runs over the positive frequency bins and $\tilde{\mu}_j(\theta)$ is the frequency-domain representation of the signal model $M$ for parameters $\theta$.
While common in the general field of time-series data analysis, the term Whittle likelihood is rarely used in the gravitational-wave literature (see, e.g. \citet{Veitch:2014wba}). We learned of the name from \citet{Thrane:2018qnx} and chose to apply it here to distinguish it from the GP likelihood, which we derive later in this text.
If data from multiple detectors are available, assuming that the noise is uncorrelated, the network log-likelihood is constructed from the sum of the log-likelihood for each detector (and this applies to both the Whittle likelihood and GP approaches).

\cref{eqn:whittle} was first derived by Peter Whittle \citep{whittle1953analysis} as an approximation to the Gaussian likelihood for a stationary time series:
\begin{align}
    \ln\likelihood\left(\data | \parameters, \model\right) =
    & -\frac{1}{2}\vec{r}(\theta)^T \Sigma^{-1}\vec{r}(\theta) -\frac{1}{2}\ln\left((2\pi)^N |\Sigma|\right)\,,
    \label{eqn:gaussian}
\end{align}
where $\vec{r}(\parameters)=\data - \mu(\vec{t}, \parameters)$ is the residual after subtracting the signal from the data, $\Sigma$ is the covariance matrix of the background noise, and $N=f_s T$ is the number of data points.
The Whittle likelihood approximation (\cref{eqn:whittle}) is advantageous compared to the full Gaussian likelihood (\cref{eqn:gaussian}) because the DFT can be replaced by the fast Fourier transform (FFT) algorithm, reducing the computational cost from $\mathcal{O}(N^3)$ to $\mathcal{O}(N \log N)$.
A recent pedagogical review \citep{rao2021reconciling} demonstrated the differences between the two likelihoods by deriving a frequency-domain representation of the Gaussian likelihood and found the difference to be of order $\mathcal{O}(1/N)$.
For typical gravitational-wave inference problems, a sampling frequency of at least a few kHz is needed to ensure the Nyquist frequency captures the signal properties during the merger and ringdown, and a duration of a few to tens of seconds is typical.
Therefore, we expect only slight differences between the Whittle and Gaussian likelihoods.
However, there are known deficiencies in the Whittle likelihood approach.
For example, recently, \citet{Talbot:2020auc} and \citet{Talbot:2021igi} have investigated the impact of the fixed PSDs and windowing of the time series, respectively.
Each of these could bias individual results at a level equivalent to other systematic uncertainties, and the effect can be magnified in population studies.

For the prior distribution of \cref{eqn:posterior}, as a general rule, the standard approach is to apply uninformative priors where possible (e.g. isotropic priors on the sky position). However, in some cases, these are informed by the output of the detection pipeline (e.g. we restrict the time of the signal to some relevant range) or by other astrophysical data (e.g. the observation of an electromagnetic counterpart may imply certain restrictions on the orientation of the source).
In this work, we apply standard prior definitions used throughout the field (see, e.g. \citet{Romero-Shaw:2020owr}).

Having defined the likelihood and priors, to estimate \cref{eqn:posterior}, either a stochastic-sampling \citep{Veitch:2014wba, Cornish:2014kda, Biwer:2018osg, Ashton:2018jfp} or grid-based \citep{Lange:2018pyp} algorithms is used. These methods are computational-intensive, typically needing $\mathcal{O}(10^8)$ or more evaluations of the likelihood and prior, with each evaluation taking $\gtrsim$\qty{1e-3}{\s} (typically, this time is dominated by the generation of the waveform and the overheads of the Whittle likelihood).
Therefore, the typical time needed to estimate the posterior distribution is about a day.
However, this cost can increase dramatically if, for example, more physically accurate waveforms are used or the signal duration requires a longer data span.
The total wall time required for analysis can be reduced by parallelisation and optimisation of the likelihood in addition to methods that reduce the cost of the likelihood and waveform evaluation.
Nevertheless, Bayesian inference remains expensive but necessary to resolve the highly-correlated structure in the posterior distribution robustly (whereby ``robustly'' we mean that repeated analyses reproduce the same result within statistical uncertainties).

\cref{eqn:whittle} has been applied with great success to every CBC signal observed by advanced-era detections \citep{GWTC3}.
However, it cannot be applied directly because, as discussed in the introduction, up to 20\% of these are contaminated by glitches, violating the fundamental assumption that the background noise is stationary and Gaussian.
To address this, the \bayeswave algorithm extends \cref{eqn:posterior}, including an additional glitch model (and associated set of model parameters).
Performing inference on this glitch model (simultaneously with the signal model), a model of the glitch is constructed and then subtracted from the data.
The standard inference process then uses the glitch-subtracted data in \cref{eqn:whittle}.
The difficulty arises if the glitch is imperfectly removed. Then, the residual power will remain in the glitch-subtracted data and bias the inference of the signal properties.
In the next section, we describe our proposed modification to the standard methodology, replacing \cref{eqn:whittle} with a GP likelihood.

\subsection{Gaussian processes}
\label{sec:gp}

A GP models a time series as a multivariate Gaussian distribution with a mean function $\mu(\vec{t}; \parameters)$ and a kernel function $k(t_m, t_n; \alpha)$ where $\alpha$ is a set of hyperparameters describing the noise process.
Under this interpretation, the log-likelihood is given by 
\begin{align}
\ln\likelihood\left(\data | \parameters, \alpha, \model\right) =
& -\frac{1}{2}\vec{r}(\theta)^T \Sigma(\alpha)^{-1}\vec{r}(\theta) -\frac{1}{2}\ln\left((2\pi)^N |\Sigma(\alpha)|\right)\,.
\label{eqn:gplikelihood}
\end{align}
Comparing the GP likelihood with \cref{eqn:gaussian}, we see that we have replaced the (fixed) covariance matrix $\Sigma$ with the covariance matrix generated by the kernel function:
\begin{equation}
    \Sigma_{mn}(\alpha) = k(t_m, t_n; \alpha)\,.
    \label{eqn:kernel}
\end{equation}

GPs are powerful because they enable us to model the noise process instead of the noise realisation.
The art of GP modelling is constructing a kernel that captures the noise process and inferring the hyper-parameters.
Simultaneously, we can infer the properties of the mean model $\mu(\vec{t}; \parameters)$, which will be marginalised with respect to the uncertainty in the GP noise model.

Direct application of \cref{eqn:gplikelihood} is infeasible for gravitational-wave signals since, as with the full Gaussian likelihood the computation cost is of order $\mathcal{O}(N^3)$.
However, if the kernel function is stationary and constructed from a mixture of exponential functions, \citet{celerite} demonstrated that the likelihood can be calculated with linear scaling by exploiting structure in the covariance matrix. The authors implement this method in the \celerite software, which we will use throughout this work.
Such kernel functions are not abstract: they
arise naturally from random processes consisting of stochastically-driven
damped harmonic oscillators. In a meaningful sense, many of the noise
processes afflicting gravitational-wave detectors can be considered a
collection of harmonic oscillators. Therefore, it is natural to hypothesise
that such fast and scalable GPs may be brought to bear on the problem of
gravitational-wave data analysis.
To our knowledge, GPs have yet been applied to this problem (though they have been applied in other settings in the field, for example to estimate model uncertainty \citep{Moore:2015sza, Doctor:2017csx, Williams:2019vub} and for interpolating posterior inferences \citep{DEmilio:2021laf}).

\subsection{The GP likelihood for gravitational-wave signals}
In this work, we will utilise the \celerite software to develop a gravitational-wave GP likelihood capable of inferring the properties of gravitational-wave signals contaminated by glitches.
We will exclusively use the Simple Harmonic Oscillator (SHO) kernel described in \citet{celerite}, which is characterised by an oscillator frequency $\omega_0$, a quality factor $Q$, and amplitude $S_0$.
The SHO kernel is stationary: it depends only on the difference $t_m - t_n$ (this is true of all kernels in \celerite).
Therefore, the stationary SHO kernel would appear to be a poor fit to model non-stationary transient glitches.
Indeed, in practice, the SHO kernel performs poorly for short-duration glitches that last for only a fraction of a second.
However, as we demonstrate later, the SHO kernel can model the underlying noise process for glitches that last an appreciable fraction of the signal duration.
We discuss this further in \cref{sec:conclusion} and suggest new avenues for applying transient GP models.

In principle, a GP model could be built that models the coloured Gaussian background noise characteristic of gravitational-wave detectors.
This could be done using a mixture of SHO terms using terms with small quality factors to model the coloured noise and SHO terms with large quality factors to model the narrow Lorentzian features (see \citet{Littenberg:2014oda} for a discussion of how \bayeswave performs a similar task).
However, for this initial study, where we are primarily focused on the ability of the GP to model glitches, we will instead utilise a technique known as \emph{pre-whitening}.

To pre-whiten the data, we first construct a PSD $P_j$ using the median Welch method \citep{welch1967}.
We then FFT the time series data to obtain $\tilde{d}$ and then apply the whitening transform:
\begin{equation}
    \hat{d}_i = \textrm{IFFT}\left(\frac{2}{N}\frac{\tilde{d}}{\sqrt{P}}\right)\,,
\end{equation}
where IFFT is the inverse FFT and the factor of $2/N$ normalises the whitened data. If the PSD properly represents the coloured Gaussian noise, then the whitened data $\hat{d}$ is zero mean with unit variance.

Our GP model is then applied to whitened data $\hat{d}$ using a fixed PSD.
To calculate the likelihood given a set of GP hyperparameters $\alpha$ and model parameters $\theta$, we apply \cref{eqn:gplikelihood} but note that the mean model $\mu(\vec{t}; \parameters)$ must also be whitened by the same PSD.

By default, our GP kernel (\cref{eqn:kernel}) then consists of a white-noise term with a single hyperparameter $\sigma$ in addition to $M$ SHO terms.
Each SHO term adds three parameters: frequency, quality factor, and amplitude.
This arrangement was developed by trialling several off-the-shelf kernels in the \celerite library.
Ultimately, we found the SHO worked the best to capture the glitch features studied in this work (as evaluated by posterior predictive checks).

When the kernel includes multiple SHO terms (i.e. $M>1$), we have a label-switching degeneracy which we find severely restricts the efficiency of stochastic sampling.
To resolve this, we apply an order-statistics prior such that the joint prior on the set of SHO log-frequencies, $\{\ln(\omega_0^{(m)})\}$ (indexed by $m$, the SHO term) is uniform, but the marginal prior is ordered.
For the remaining SHO hyperparameters, we apply uniform priors on the logarithm of the values.

\section{Results}
\label{sec:results}

We apply the GP method to three case studies to demonstrate its capability and explore the limitations of the current implementation.
The first two studies investigate a binary black hole (BBH) signal overlapping examples of scattered-light glitches.
These glitches were common during O3 and have been linked to periods of increased ground motion around the detectors which in turn cause light scattering from the high-power laser beam \citep{LIGO:2021ppb}.
Two distinct scattered light glitches are known: slow-scattering \citep{LIGO:2020zwl} and fast-scattering \citep{Soni:2021cjy}; we discuss each in turn in Case Studies A and B.
The final case study demonstrates the use of the GP method as a tool to perform time-domain tests of General Relativity.
We perform an Inspiral-Merger-Ringdown test on the first observed BBH system, GW150914, separating the signal in time rather than frequency.
Our results are consistent with previous analyses but demonstrate a resolution to spectral leakage issues.
For all three case studies, we perform Bayesian inference using the \texttt{Bilby} Bayesian inference library \citep{Ashton:2018jfp} and the \texttt{Bilby-MCMC} sampler optimized for BBH data analyses \citep{Ashton:2021anp}.

\subsection{Case study A: A slow-scattering glitch}
\label{sec:caseA}

We begin with a single-detector study of one of the two types of scattered-light glitches: a slow-scattering glitch. Slow-scattering glitches are characterised by long-duration $\mathcal{O}(1-4)$~s stacked arches when viewed in a spectrogram. Multiple arches are formed due to multiple reflections of the scattered light \citep{LIGO:2021ppb}.

\begin{figure*}
    \centering
    \includegraphics{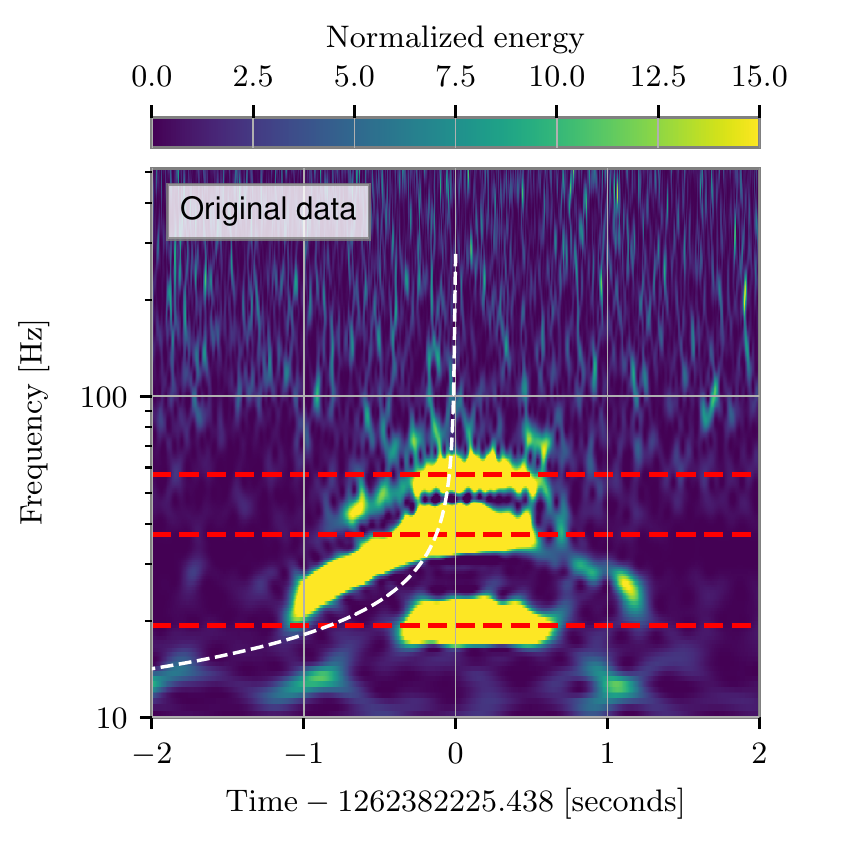}
    \includegraphics{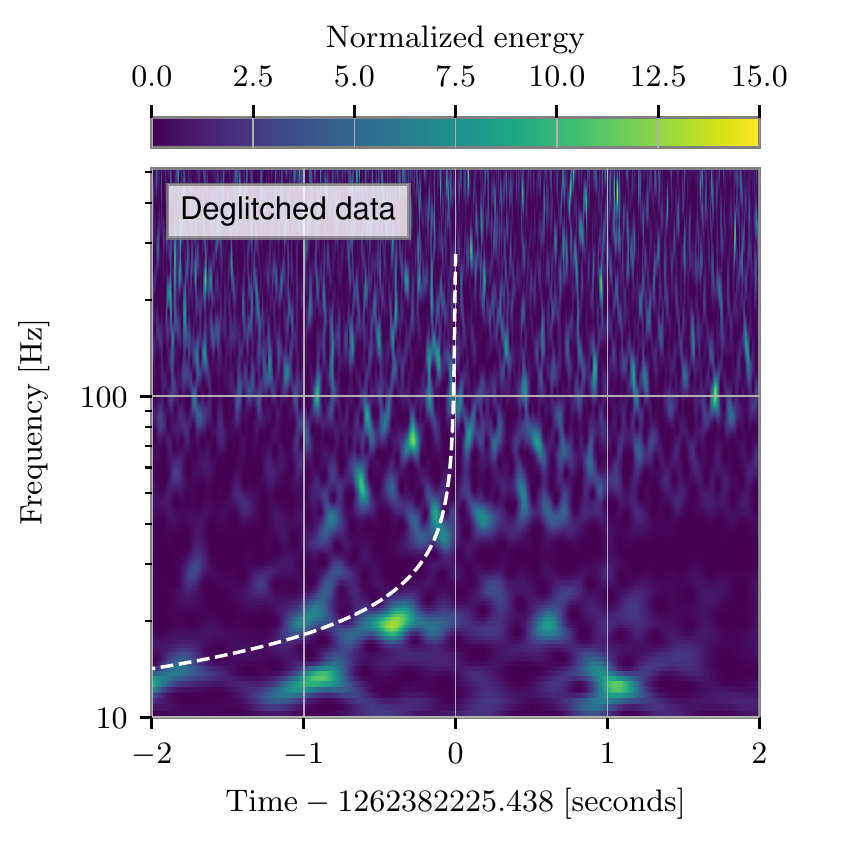}
    \caption{Spectrograms of the slow-scattering glitch analysed in case study A plotted alongside the frequency track of the simulated signal. In the left-hand panel, we show the original strain data, whereas the right-hand figure shows the strain data after \bayeswave glitch subtraction is applied. Note that the glitch subtraction is applied \emph{before} the simulated signal is added.
    We also show dashed horizontal lines in the left-hand panel highlighting the median SHO frequency inferred in the GP analysis.}
    \label{fig:SS_data}
\end{figure*}

We take a representative example identified by \gravityspy \citep{zevin_2017, Glanzer:2022avx} in the LIGO Livingston interferometer during the O3 observing run.
In the left-hand panel of \cref{fig:SS_data}, we plot a spectrogram of the data illustrating three distinct scattering arches between 10 and \qty{70}{\Hz}.
Overlaid on this figure is the time-frequency track of the simulated signal that we will add to the data (it is not included in the spectrogram to avoid confusion with the glitch itself).
Before adding the simulated signal, we apply the \bayeswave deglitching routine using standard settings.
The resulting spectrogram (right-hand panel of \cref{fig:SS_data}) demonstrates some residual glitch power below \qty{20}{\Hz} which we were unable to remove, but this residual power does not fall along the frequency-time track of the injected signal.

In contrast to the analyses of overlapping signals and glitches in real data, the deglitching process in this case study is applied in the absence of a signal.
Therefore, it is not representative of a real analysis but does represent an ideal of the deglitching process: the glitch can be characterised and subtracted without signal confusion.
However, in practice, this will not usually be possible unless an auxiliary channel provides a means to model the glitch without using the strain data. Instead, for typical cases, which use \bayeswave to deglitching the data, the signal and glitch must be modelled simultaneously. This leaves open the possibility for erroneous signal removal. Nevertheless, it is a good demonstration of how the GP method compares against the ideal deglitching scenario.

Having both the original data and the deglitched data, we can perform a direct comparison of three different approaches. The first, which we label ``BW+WL'' applies the standard Whittle likelihood to data which has been deglitched using \bayeswave. Therefore, BW+WL denotes the method used by the LVK collaborations to date.
The second, which we label ``GP'', is to apply the GP method directly to the original data, modelling the glitch and signal simultaneously. 
Finally, we label analyses where the Whittle likelihood is applied directly to the original data ``WL''; these analyses illustrate the extent to which the glitch biases inferences in the absence of a mitigation technique.
In validating the method, we did also analyse the deglitched data using the GP approach. As expected, we found near identical results to the BW+WL approach since the GP simply sets the hyperparameters of the SHO such that its effect is negligible, and then one recovers the standard Whittle likelihood.

In each analysis, taking either the original or deglitched data, we add a simulated equal-mass non-spinning BBH signal generated using the time-domain \imrt waveform approximant \citep{Estelles:2020twz}. The process of generating this analysis (i.e. obtaining the strain data and adding the simulated signal) is independent of the analysis method, ensuring a straightforward comparison can be made.
The signal track of the simulated signal is shown as a dashed curve in \cref{fig:SS_data}.

We then analyse the original data (including the simulated signal) using the WL and GP approach and the deglitched data using the BW+WL approach.
All analyses are performed on data from a single detector, though they naturally scale to multiple detectors by adding together the log-likelihoods assuming independent noise.
For the comparative case studies (A and B), we infer the signal properties using the \imrt waveform, which assumes the spin of both black holes is aligned along their angular momentum and we a priori fix the sky position of the source to the simulated values.
We apply these simplifications because they reduce the overall computational cost by nearly an order of magnitude with minimal impact on the comparison;
ultimately, we are interested in the ability of the GP model to separate the signal from the noise.
However, we do note that in this feasibility study, we have therefore not examined the interaction of a signal with misaligned spins, which will exhibit precession effects, with the GP method.
Finally, for all analyses, we generate a PSD by applying the median Welch method to off-source data.
For the WL and BW+WL analysis, the PSD is used directly in the likelihood, while for the GP method, the PSD is used to whiten the data and the predicted model.

\begin{figure*}
    \centering
    \includegraphics{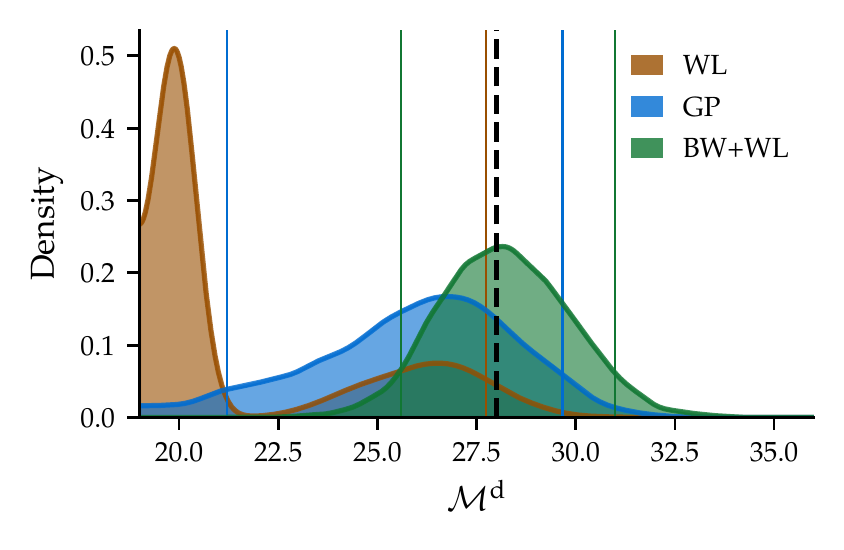}
    \includegraphics{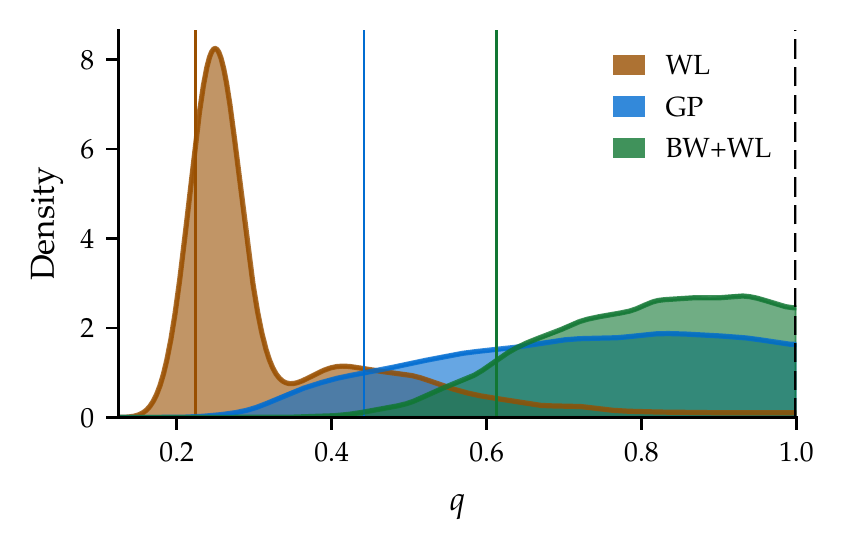}
    \includegraphics{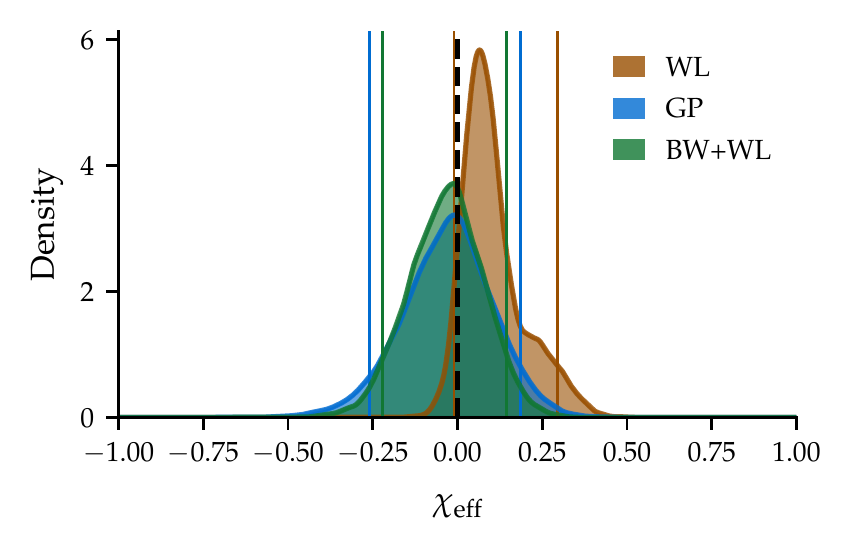}
    \includegraphics{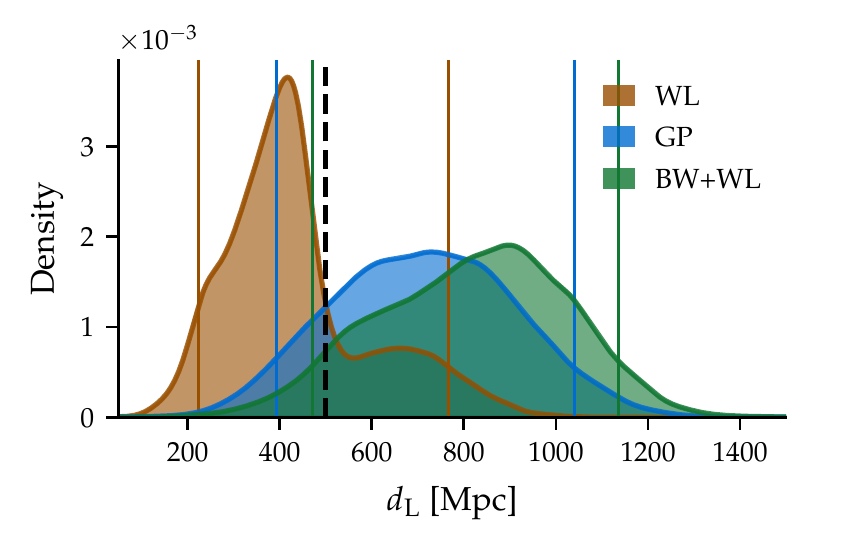}
    \caption{The posterior distribution for the detector-frame chirp mass \chirpmass, mass ratio \massratio, effective spin \chieff, and luminosity distance \luminositydistance for case study A. In each figure, we compare the WL applied to the original data (WL), the GP applied to the original data (GP), and the WL applied to the deglitched data (BW+WL). The simulated signal value is given as vertical dashed lines while 90\% credible intervals are given as coloured vertical lines for each distribution.}
    \label{fig:SS_posterior}
\end{figure*}

In \cref{fig:SS_posterior}, we compare kernel density estimates of four source parameters inferred using the WL, GP, and BW+WL approaches.
We choose to compare the detector-frame\footnote{For gravitational-wave radiation from sources at cosmological distances, the signals are red-shifted. As a result, it is not possible to measure the source-frame mass directly, but instead, we measure the detector-frame mass. The former can be inferred by supplying a cosmological model to estimate the redshift $z$ from the measured luminosity distance and then scaling the detector-frame mass by $1 / (1 + z)$ \citep{Cutler:1994ys}.}
chirp mass \citep{Finn:1992xs}:
\begin{equation}
    \chirpmass = \frac{(m_1^{\rm d}m_2^{\rm d})^{3/5}}{(m_1^{\rm d} + m_2^{\rm d})^{1/5}}\,,
\end{equation}
(where $m_1^{\rm d}$ and $m_2^{\rm d}$ are the detector-frame mass of the two black holes),
the mass ratio: 
\begin{equation}
    q = m_2^{\rm d}/{m_1^{\rm d}}\,,
\end{equation}
the effective spin \citep{Santamaria:2010yb, Ajith:2009fz}:
\begin{equation}
    \chieff = \frac{\chi_1 + q \chi_2}{1 + q}\,,
\end{equation}
where $\chi_1$ and $\chi_2$ are the aligned spin components of the two black holes, and the luminosity distance, $d_L$.

The kernel used in the GP analysis presented in \cref{fig:SS_posterior} is formed from the addition of a white noise component (characterised by its standard deviation) and $M=3$ SHO terms (characterised by the frequency, quality and amplitude factors) resulting in a total of 10 noise hyperparameters.
The choice of three SHO terms and the prior applied to their hyperparameters was reached by manual tuning:
using a greater number of SHO terms, their effect was negligible, while using fewer terms, the source-parameter inferences were biased.
In this instance, the optimal number of three was also suggested by the observation of the spectrogram \cref{fig:SS_data} where we observe three distinct features.
It is then unsurprising that the median SHO frequency inferred from the GP analysis coincides nicely with the three arches in \cref{fig:SS_data} (see horizontal dashed lines).
But, this also illustrates a clear deficiency of the GP model: we assume the GP kernel is stationary, but \cref{fig:SS_data} suggests that the frequency is, in fact, changing with time.
Moreover, models of light scattering \citep{Longo:2020onu} have previously demonstrated the frequency is changing with time.
Therefore, we predict that non-stationary kernels may produce better results for transient glitches with a short duration relative to the length of data being studied.
In future, we could investigate the use of the Bayesian evidence to select the optimal number or a Reversible-Jump MCMC able to optimize over the number.
Finally, we add that for all three SHO terms, the median inferred quality factor was $\sim 30$.

Let us now consider each of the analyses in \cref{fig:SS_posterior} in turn.

First, for the WL analysis, which applies no glitch mitigation, the inferred source parameters are severely biased: this is clearly demonstrated by the measured chirp mass and mass ratio where the peak of the distribution lies well away from the simulation value. This fact is well known (see, e.g., \citep{Pankow:2018qpo}).
In recent work, we demonstrated \citep{Ashton:2021tvz} that analysing scattering glitches alone with a BBH model resulted in inferred mass ratios of $\sim 0.2$.
That the posterior for the mass ratio under the WL analysis in \cref{fig:SS_data} peaks around this value is therefore consistent with the interpretation that the WL analysis effectively \emph{misses} the simulated signal and tries to fit the BBH model to the glitch instead.

Second, the BW+WL approach is successful: all four parameters are recovered within the 90\% interval.
This observation is entirely unsurprising, of course: the success of the BW+WL approach is already well known in the literature, and \cref{fig:SS_data} demonstrates visually that the \bayeswave algorithm successfully deglitches the data.

\begin{figure}
    \centering
    \includegraphics{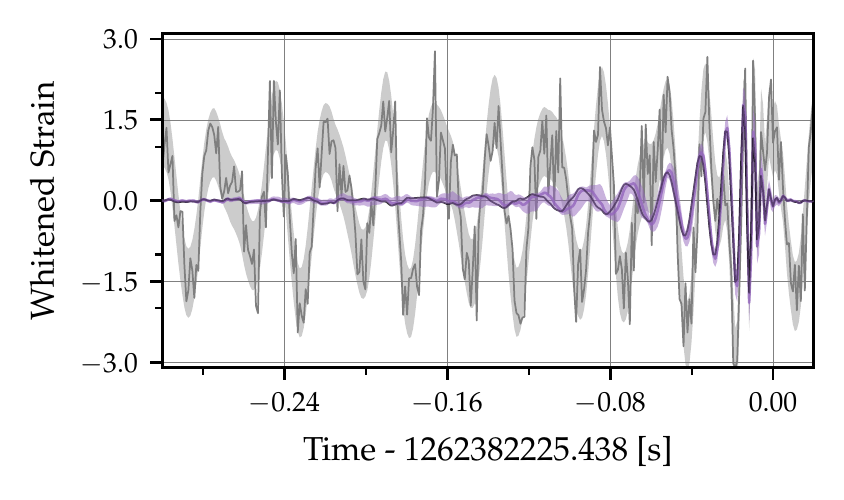}
    \includegraphics{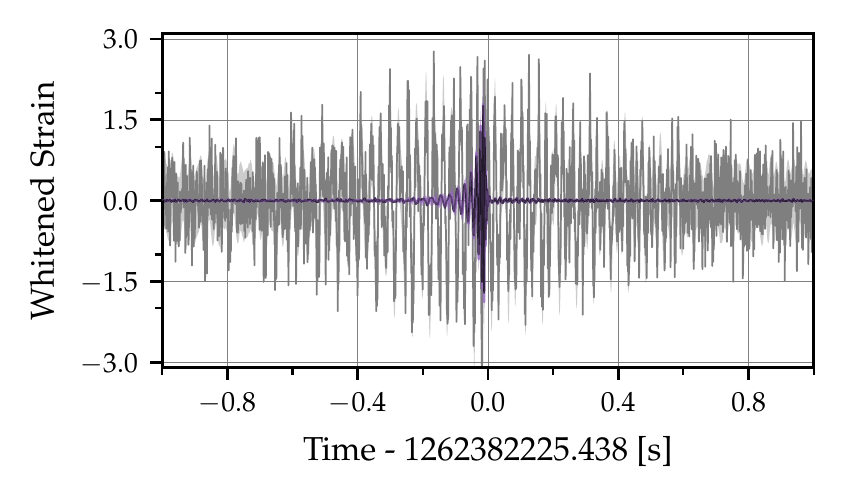}
    \caption{
    A posterior predictive plot for the slow-scattering glitch (case study A).
    In a solid grey curve, we plot the whitened strain data from LIGO Livingston; this includes the simulated signal added to the data, which we also plot as a solid black line (after whitening).
    In purple, we plot the posterior model prediction: a solid curve is generated from a median estimate of the source parameters, while the purple-shaded region indicates the 90\% credible interval for the model prediction (generated by drawing repeated predictions from the model from the posterior distribution).
    Finally, in a solid grey band, we plot the 90\% credible interval for the inferred GP noise by repeatedly drawing predictions from the GP kernel with a zero mean.
    The upper panel provides a close look at the features of the signal near to the merger, the lower panel includes a wider span of data illustrating the features of the glitch (cf. \cref{fig:SS_data}).
    }
    \label{fig:SS_pp}
\end{figure}

Finally, like the BW+WL approach, the GP approach also successfully recovers the simulated values within the 90\% credible intervals.
This indicates it has successfully modelled the signal as part of the mean model and the glitch as part of the noise.
To further check this, in \cref{fig:SS_pp}, we create a posterior predictive test showing the modelled glitch and noise alongside the simulated signal.
The posterior predictive plot demonstrates the difficult task the GP analysis has completed: to pick out the signal from the highly non-Gaussian noise.

Comparing the GP and BW+WL approaches in \cref{fig:SS_posterior}, we see that the GP credible interval is, in all cases, wider than that of the BW+WL method.
Taken at face value, this suggests the BW+WL method is preferable since it constrains the source parameter with greater precision.
However, we remind the reader that the BW+WL approach is operating under perfect conditions, where the glitch is modelled in the absence of the signal.
In practise, this will not be the case.
On the other hand, the GP approach results are marginalised over the uncertainty induced by the glitch, which increases the uncertainty.
Therefore, the increase in uncertainty represents exactly the goal of the GP analysis: to jointly constrain both the astrophysical signal and glitch and circumvent the need for deglitching which intrinsically neglects uncertainty.
However, we do also remark that the BW+WL and GP glitch models are fundamentally different and, it could be the case that the \bayeswave model is in some way better than the SHO model.
To test this, we will need to compare a full joint analysis by \bayeswave of the glitch and BBH signal as done in \citet{Hourihane:2022doe}.
Hence, for the time being, we can conclude that the GP process is able to model the glitch and signal simultaneously, but further comparisons are needed to understand the relative performance of the two approaches.

\subsection{Case Study B: An O3 fast-scattering glitch}
\label{sec:caseB}

\begin{figure}
    \centering
    \includegraphics{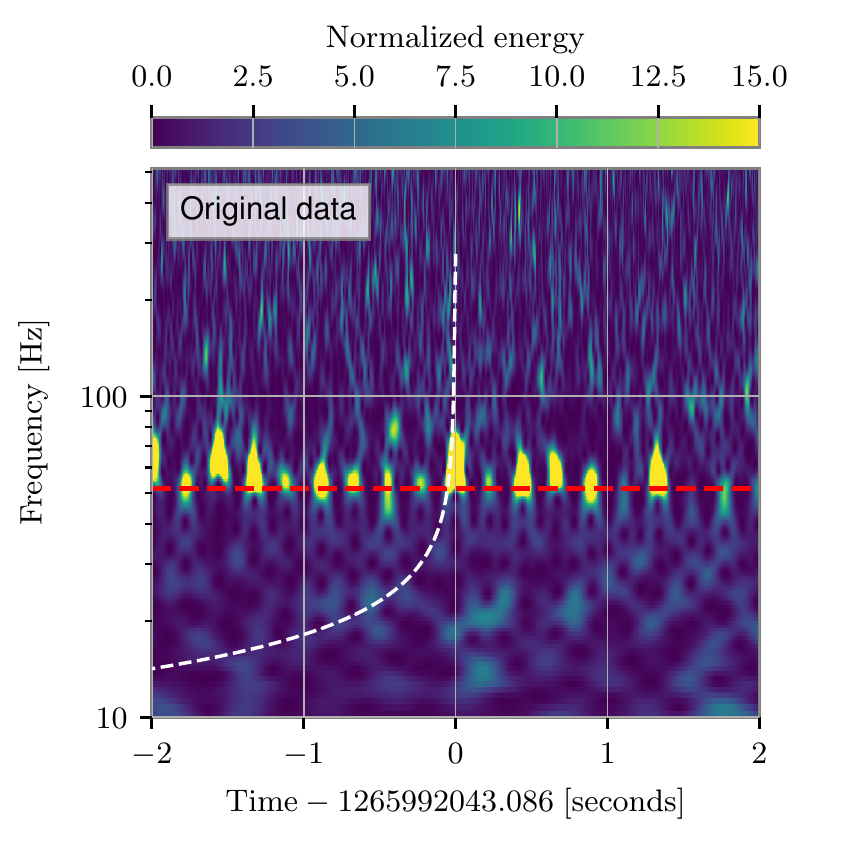}
    \caption{Spectrogram of the fast-scattering glitch analysed in case study B plotted alongside the frequency track of the simulated signal.
    We also show dashed horizontal lines highlighting the median SHO frequency inferred in the GP analysis.}
    \label{fig:FS_data}
\end{figure}

In our second case study, again using single-detector data, we add a simulated signal to data containing a fast-scattering glitch.
The original data is shown in \cref{fig:FS_data} and illustrates the typical behaviour of this glitch class: a series of short-duration artefacts centred around (in this instance) a frequency of $\sim$\qty{60}{\Hz}.
We did try to excise the glitch from the original data.
However, we found it difficult to find adequate \bayeswave configuration which removed the glitch (though \citet{Hourihane:2022doe} do report successful modelling of other fast scattering glitches).
For this reason, we cannot perform the BW+WL analysis.
Therefore, we instead perform a repeated WL-offset analysis but shift the simulated signal time by $+$\qty{20}{\s}.
During this time, the data does not contain any transient noise artefacts (though we note the estimated PSD will).

To the original data, we add a simulated signal (identical to that of case study A); the time-frequency evolution of this signal is shown in \cref{fig:FS_data} and then perform three analyses: the WL, GP and WL-offset analysis.
For the GP model, we find that a kernel consisting of a single SHO term in addition to a white-Gaussian noise term was sufficient to model the background noise and extract the signal properties.
In \cref{fig:FS_data}, we plot the median inferred frequency of the SHO term as a horizontal bar demonstrating that it captures the average frequency of the repeating transient artefacts.
The quality factor the SHO term inferred from the data was $\sim 10$.

\begin{figure*}
    \centering
    \includegraphics{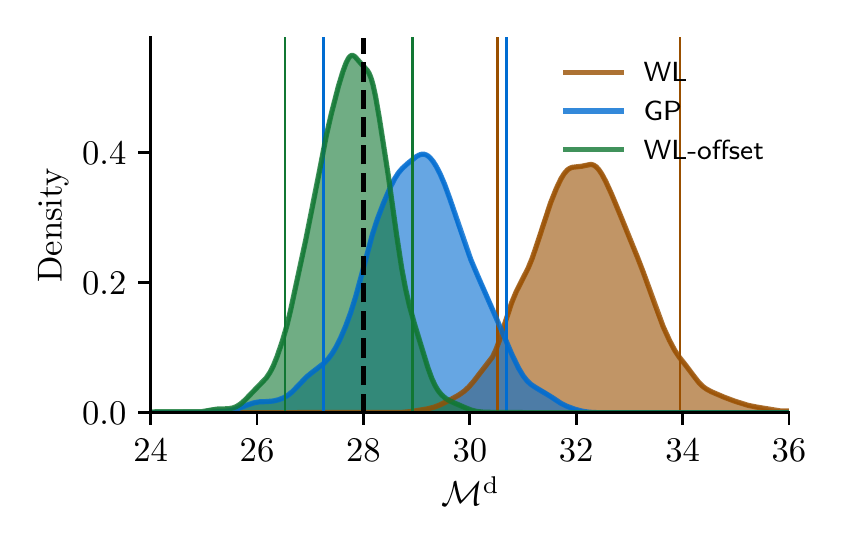}
    \includegraphics{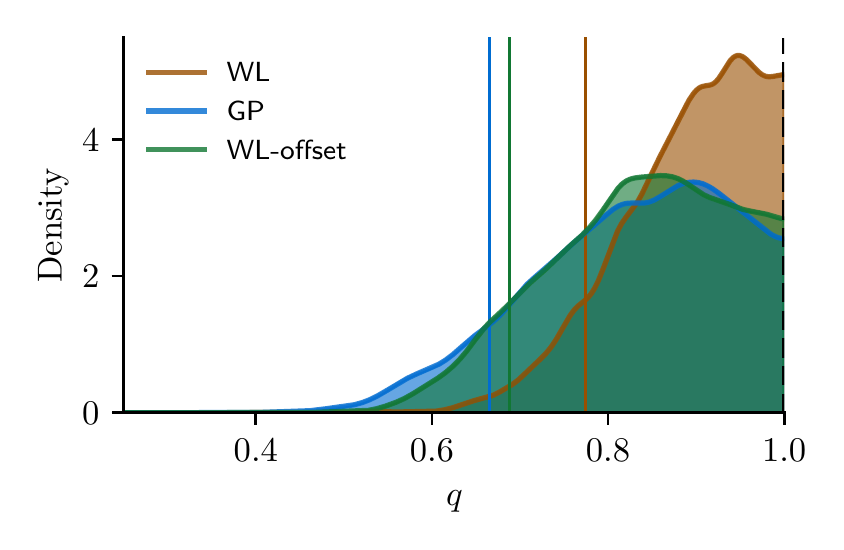}
    \includegraphics{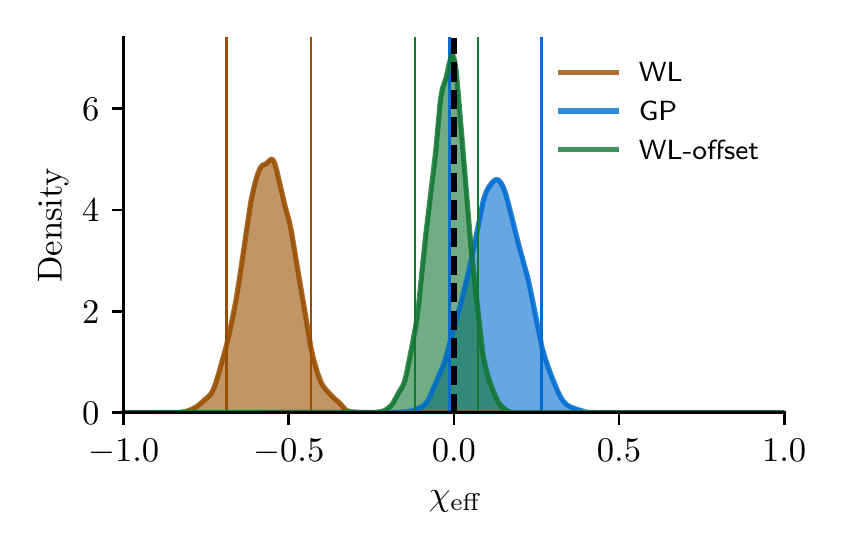}
    \includegraphics{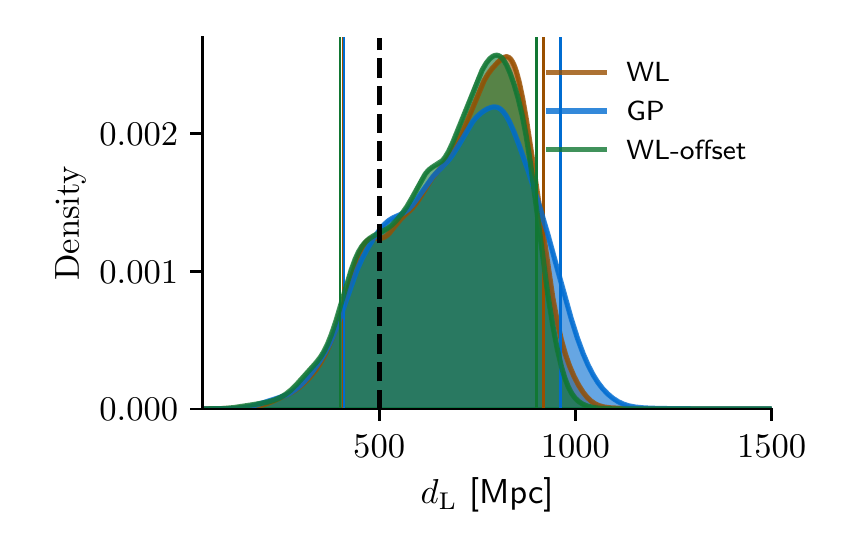}
    \caption{The posterior distribution for the detector-frame chirp mass \chirpmass, mass ratio \massratio, effective spin \chieff, and luminosity distance \luminositydistance for case study B. In each figure, we compare the WL applied to the original data (WL), the GP applied to the original data (GP), and the WL applied to an offset data segment without a glitch (WL-offset).
    The simulated signal value is given as vertical dashed lines while 90\% credible intervals are given as coloured vertical lines for each distribution.}
    \label{fig:FS_posterior}
\end{figure*}

In \cref{fig:FS_posterior}, we plot kernel density estimates of the posterior from the three analyses.
Once again, the WL analysis demonstrates the severe bias by an analysis which is unable to model the glitch.
Meanwhile, the GP and WL-offset analyses both recover the signal within their 90\% credible intervals.
However, the WL-offset analysis is better constrained.
This is to be expected for the same reasons discussed in the context of case study A.
The difference is even more severe in this case.
We expect that this is due to the particularity of the glitch under study.

\begin{figure}
    \centering
    \includegraphics{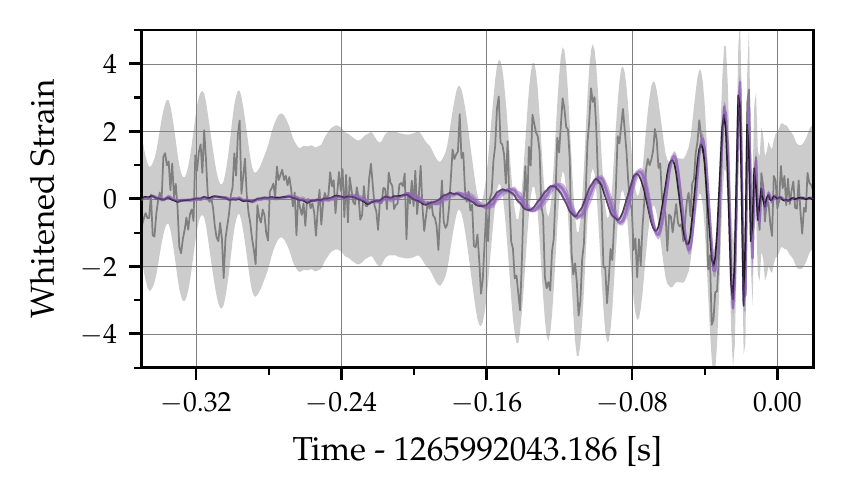}
    \caption{
    A posterior predictive plot for the fast-scattering glitch (case study B).
    In a solid grey curve, we plot the whitened strain data from LIGO Livingston; this includes the simulated signal added to the data, which we also plot as a solid black line (after whitening).
    In purple, we plot the posterior model prediction: a solid curve is generated from a median estimate of the source parameters, while the purple-shaded region indicates the 90\% credible interval for the model prediction (generated by drawing repeated predictions from the model from the posterior distribution).
    Finally, in a solid grey band, we plot the 90\% credible interval for the inferred GP noise by repeatedly drawing predictions from the GP kernel with a zero mean.
    }
    \label{fig:FS_pp}
\end{figure}

In \cref{fig:FS_pp}, we create a posterior-predictive plot which demonstrates the overlap of the signal and glitch in the inspiral phase.
From this plot, we can understand why the WL analysis, which assumes only Gaussian noise and a BBH signal, is so severely biased by the oscillatory features of the glitch.
Furthermore, we can understand why the WL-offset analysis is far better constrained than the GP analysis.
For the GP analysis, the inspiral is severely contaminated by the glitch, effectively reducing the SNR and hence resulting in a wider posterior compared to the WL-offset analysis where the inspiral is uncontaminated by the glitch.

\subsection{Case Study C: Enabling time-domains tests of General Relativity: Inspiral-Merger-Ringdown}
\label{sec:caseC}

In our final case study, we discuss how the GP approach can enable time-domain tests of General Relativity (GR) which are robust to glitches.
The violent final stages of a binary black hole collision offer a unique opportunity to probe the limits of GR in the strong-field regime (see \citet{LIGOScientific:2021sio} for the latest results).
There are a variety of proposed tests, including theory-independent consistency tests (looking for generic fails of the predicted GR model) and tests which search for explicit predictions (see, e.g. \citet{Yunes:2013dva, Yagi:2016jml} for reviews).
However, most of these tests rely on the frequency-domain data analysis approach developed for source parameter estimation.
While this is not inherently problematic, there is a subset of cases where the frequency-domain approach forces awkward workarounds to implement what are intrinsically time-domain tests.

An example of this is the so-called Inspiral-Merger-Ringdown (IMR) consistency test which, conceptually, breaks the data in the time domain into two segments, the inspiral and merger-ringdown \citep{LIGOScientific:2016lio, Ghosh:2016qgn, Ghosh:2017gfp}.
The IMR test then checks for consistency between the mass and spin of the remnant Kerr black hole, as predicted by the inspiral and the merger-ringdown data separately.
The difficulty arises in that the analysis is performed in the frequency domain, not the time domain.
The separation of inspiral and merger-ringdown data is practically performed by choosing a cutoff frequency.
Data below the cutoff is assumed to arise from the inspiral, while data above the cutoff is taken to be of the merger-ringdown.
However, this practical choice can result in spectral leakage if the inspiral signal contains power at frequencies above the cutoff or the merger-ringdown signal contains power below the cutoff.
\citet{Ghosh:2017gfp} demonstrate that if the cutoff frequency is chosen to be that of the innermost stable circular orbit (ISCO) of the final Kerr back hole, then the amount of leakage is small.
But, we point out that this validation was performed on a circular aligned-spin system.
For precessing systems (and systems including the effects of eccentricity), the frequency evolution during the inspiral can be non-monotonic, and the signal can contain frequency content above the ISCO, which would result in spectral leakage between the inspiral and merger-remnant.
Spectral leakage is of concern because, if it occurs at a significant level, it violates the premise of the test and enforces an automatic consistency.
Moreover, even if the level of spectral leakage is small, the practical implementation of the IMR consistency test (splitting the signal in the frequency domain) is at odds with the conceptual idea of splitting the signal in the time domain.

A second example where a similar difficulty is faced is that of the ringdown-only analysis \citep{LIGOScientific:2016lio, Isi:2019aib, Carullo:2019flw, Capano:2021etf}, where data arising from the ringdown alone is analysed for consistency with the theoretical predictions of GR.
Such tests face a similar difficulty to the IMR tests: they need to take data only from the ringdown, ensuring there is no contamination from the inspiral.
In this case, a simple frequency cut will not suffice.
The two dominant approaches developed to solve the problem are ``gating and in-painting'' \citep{Capano:2021etf} where the inspiral is replaced such that the over-whitened data is 0 in the region of interest and a time-domain likelihood \citep{Carullo:2019flw, Isi:2019aib} which is able to analyse only the merger-ringdown.

\begin{figure}
    \centering
    \includegraphics{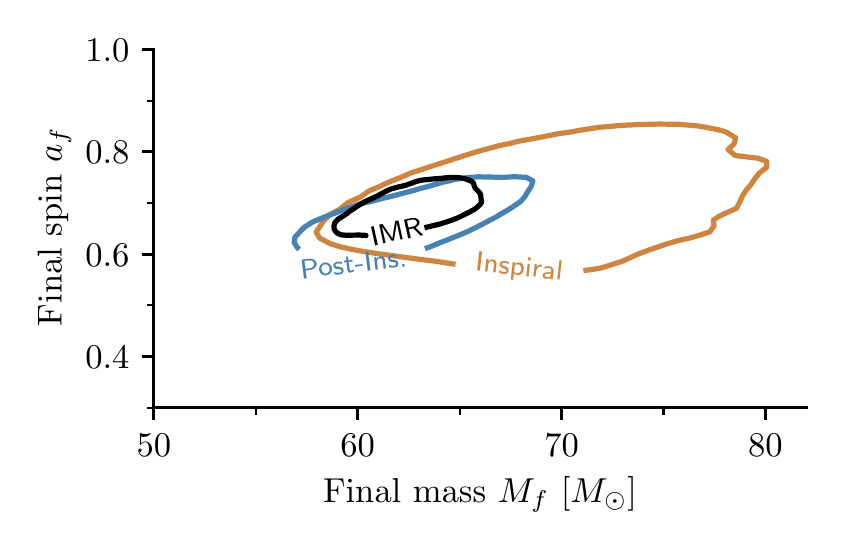}
    \includegraphics{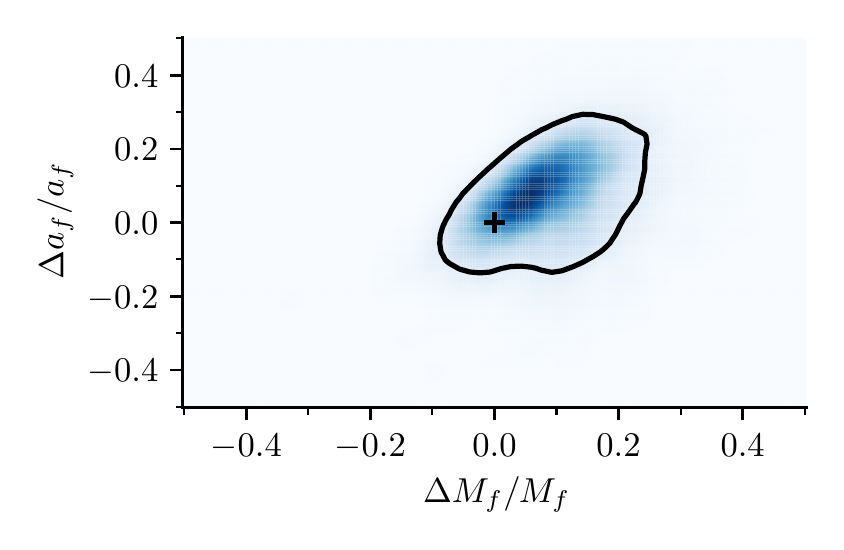}
    \caption{The 90\% credible interval of the posterior distribution for final mass and spin of the remnant black hole inferred from the GW150914 signal (top panel) and the fractional deviation parameters (bottom panel) calculated from Eq. (3) of \citet{LIGOScientific:2021sio}.
    In the top panel, the ``IMR'' curve refers to the analysis of the entire \qty{4}{\s} segment of data surrounding the event, ``Inspiral'' to the \qty{2}{\s} of data before the trigger time, and ``Post-Ins.'' to the \qty{2}{\s} of data after the trigger time.
    In the bottom panel, the black contour marks the 90\% credible interval while the ``+'' symbol marks the prediction of GR.
    }
    \label{fig:IMR}
\end{figure}

Time-domain ringdown analyses demonstrate the general-purpose applicability of time-domain approaches to testing GR.
That our GP approach is itself time-domain by construction motivates us to demonstrate how it can be readily applied to resolving the spectral leakage problem for IMR tests.
To this end, we perform a reanalysis of the first observed binary black hole merger, GW150914 \citep{LIGOScientific:2016aoc}.
Taking open data from the Gravitational-Wave Open Science Centre \citep{LIGOScientific:2019lzm}, we analyse the data from LIGO Hanford and Livingston using the precessing \imrtp \citep{Estelles:2020osj}.
We use standard prior configurations (isotropic priors on the spin, uniform in the component masses, and uninformative on the position and orientation), noting that unlike case studies A and B, this analysis makes no assumptions about the spin or sky location.
We first construct a PSD using the median Welch method from \qty{124}{\s} of data prior to the signal.
We then analyse the full signal (labelled IMR), comprising \qty{4}{\s} of data centred on the GPS trigger time $t_0=1126259462.418$, the inspiral signal comprising \qty{2}{\s} of data before $t_0$, and the merger-ringdown (labelled Post-Ins.) comprising \qty{2}{\s} of data after $t_0$.
The posterior distributions for the final mass and spin inferred from these three analyses are then presented in the upper panel of \cref{fig:IMR}.
Next, we apply the procedure described in \citet{Ghosh:2017gfp}, (see Eqn 5), to calculate the fractional difference between the inspiral and post-inspiral analyses normalized to their average and plot the resulting distribution in the lower panel of \cref{fig:IMR}.

\begin{figure}
    \centering
    \includegraphics{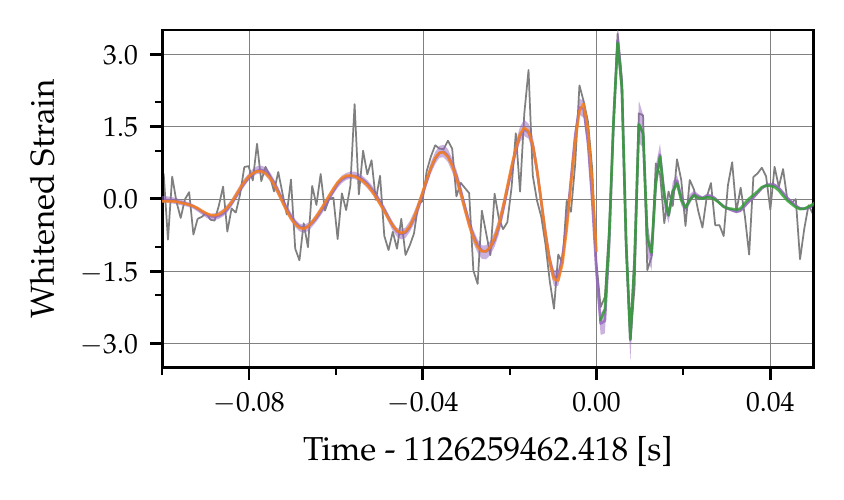}
    \includegraphics{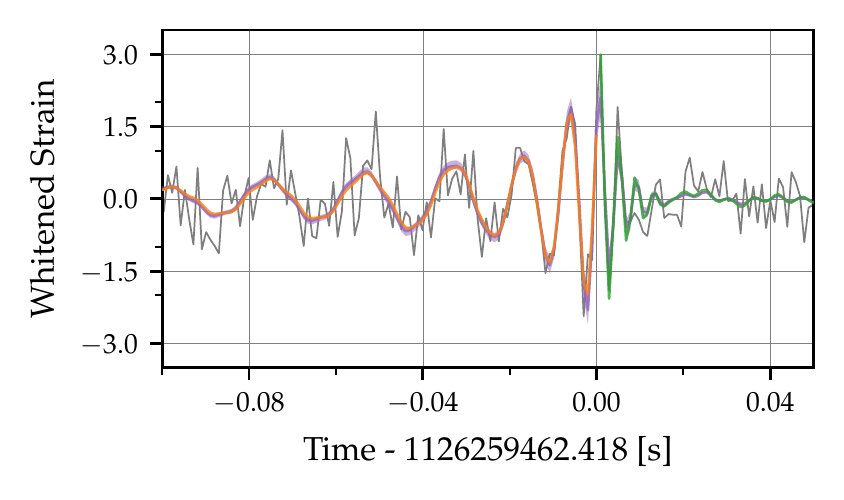}
    \caption{Comparison of the whitened strain data from LIGO Hanford (top panel) and LIGO Livingston (bottom panel), with the posterior plot from the three analyses performed in the IMR test.
    In orange (green), we plot the waveform of the median posterior inferred from the inspiral (post-inspiral) data, while in purple, we plot the 90\% credible interval
    of waveforms inferred from the full IMR analysis.}
    \label{fig:IMR_waveforms}
\end{figure}

Our results agree with the original analyses \citep{LIGOScientific:2016lio}: we find strong consistency between the inspiral and merger-ringdown predictions.
Moreover, the prediction of GR falls comfortably within the 90\% credible interval.
However, we do note that our 90\% credible interval for the fractional deviation is smaller than that of the original analysis and our post-inspiral inference is similarly better constrained.
In absence of the differing methodology, there are already several possible causes for this, including the use of a different waveform model and PSD.
However, we anticipate that the most significant differences arise due to the construction of the time-domain cutoff.
Namely, we cut the data based on the trigger time $t_0$ calculated from a full IMR analysis. However, this value refers to the peak of the 2-2 mode as measured at the Earth's centre.
Meanwhile, the light-travel-time delay means that the transition between merger and ringdown (itself a fuzzy concept) occurs at different times for the data from the two detectors.
This fact can clearly be seen in \cref{fig:IMR_waveforms} where we plot the data and the posterior predictive whitened waveforms from the IMR, Inspiral, and Post-Ins analyses.
Here, one can see that the time-domain cutoff between Inspiral and Post-Ins occurs earlier in Hanford (relative to the overall waveform morphology) than in Livingston.
A straightforward improvement would therefore be to define a detector-dependent cutoff time.
However, like with the frequency-domain analysis, the choice of cutoff is in some sense arbitrary and will invariably produce different results.
But crucially, for our time-domain analyses, any choice of cutoff is still a valid test of GR since we cleanly separate the data without the risk of spectral leakage.
Therefore, for this presentation, we choose to keep our simplified analysis fixing the cutoff time universally.
In future work, we plan to investigate detector-dependent cutoff times by choosing a fixed phase of the waveform and correcting for the light travel time delay.
Another interesting possibility opened up by the time-domain nature of our approach is to extend the number of data segments beyond just two and study consistency between them.
Such a multiple-segment IMR test would, in spirit, begin to look very much like the $\chi^2$-veto \citep{Allen:2004gu} used by search algorithms to identify glitches.

Since GW150914 was not contaminated by glitches, we did not include any SHO terms in the GP kernel.
In testing, we found that adding SHO terms did not change the results: we found near-identical inferences about the signal parameters, while the posterior distributions for the glitch SHO terms were uninformative except for setting a limit on the maximum amplitude.
In a sense, adding additional terms in the GP kernel when fitting to a signal that is thought to be uncontaminated amounts to an automated residual test of GR (an analysis where the best-fit signal is subtracted from the data and the residual is searched for excess power).
If the GP term were found to have non-zero power and one was certain the signal was not contaminated by a glitch, this would suggest that the signal model was not able to fit some features of the data.
Such a result could suggest the presence of an unmodelled part of the signal.

It is worthwhile pointing out that, since we do not include any SHO term, this case study bears parallels with extensions of the time-domain ringdown analyses to also model the inspiral and merger.
However, our new method can be easily extended to also include SHO terms to model any contaminating glitches: this could be crucial to future detections to verify that signal power is arising from an astrophysical source and not terrestrial contamination.
In future work, we will explore how the additional advantages of the GP approach  are leveraged. For example, using the GP to explicitly model the coloured Gaussian noise (so that pre-whitening is not needed) and performing time-domain GR tests on signals contaminated by glitches.

\section{Conclusion and Discussion}
\label{sec:conclusion}

We have presented a feasibility study for a novel approach to analysing transient signals observed by ground-based gravitational-wave detectors using Gaussian processes.
This new method fundamentally differs from the traditional approach in that it operates in the time domain and is able to model the physical processes generating Gaussian and non-Gaussian noise alongside an astrophysical signal.
Our first two case studies demonstrate that the true source parameters of a signal contaminated by scattered light glitches can be recovered.
With 20\% of observed signals observed by current generation ground-based detectors contaminated by signals, this marks an important step towards enabling joint analysis of glitches and signals and a ready alternative to explicit glitch modelling.
Then, in our final case study, we demonstrate that the Gaussian process approach can be easily leveraged to perform tests of General Relativity in the time domain, avoiding the spectral leakage problem inherent in the frequency-domain Inspiral-Merger-Ringdown tests.

That the Gaussian process approach models the underlying physical process generating the noise, rather than the noise itself, sets it in contrast to algorithms which model the glitches directly, like \bayeswave.
For example, in analysing the fast scatting glitch shown in \cref{fig:FS_data}, the Gaussian process approach uses just three parameters: the frequency, amplitude, and quality factor of the simple harmonic oscillator kernel (cf. \citet{celerite}).
By contrast, \bayeswave would need to construct a sine-Gaussian wavelet for each time-separated burst of power.
Thus, a GP model can greatly reduce the parameter space size compared to approaches that try to model the glitches directly.

However, the case studies discussed herein amount to only a demonstration of feasibility.
There is significant work yet to be done to realise the full potential of the GP approach.

First, we need to extend the scope of the kernel construction.
As pointed out in \cref{sec:results}, we have limited our analysis to the stationary simple harmonic oscillator kernel.
But, glitches are a non-stationary process.
Therefore, we plan to investigate the use of non-stationary kernels and alternative kernel types, which may better model other glitch classes and enable the analyses of short-duration glitches (relative to the signal duration).
In addition, we have applied the pre-whitening approach to circumvent the need to model the stationary coloured Gaussian noise.
In future work, we plan to study how to construct a full kernel which can model the coloured noise in addition to transient glitches.

Second, we need to improve the computational and sampling efficiency of our new approach.
We can demonstrate the problem by using the analyses presented for the slow-scatting glitch in Case Study A, which required the most complex Gaussian process kernel consisting of three simple harmonic oscillator terms.
Averaging over the 10 MCMC chains used in the analysis, the per-likelihood evaluation time of the Gaussian process was \qty{14\pm2}{\milli\s} while it was \qty{10\pm1}{\milli\s} for the Whittle likelihood analysis.
This modest additional cost illustrates the remarkable efficiency of the \celerite software, without which the Gaussian process would take orders of magnitude more time per evaluation.
However, the Whittle likelihood analyses are able to utilise explicitly marginalised likelihoods (see \citep{Thrane:2018qnx}) and do not model the glitch and signal simultaneously.
Thus, the Whittle likelihood analyses are exploring a smaller parameter space and hence require fewer likelihood evaluations.
In total, the GP analyses required 33 million evaluations of the likelihood to produce 200 independent samples per chain, while the Whittle likelihood analyses required just 0.6 million for the same number of samples. 
As a result, the Whittle likelihood analyses took over an order of magnitude fewer resources than the GP analysis.
For an apples-to-apples comparison, we should also include the computational cost of the \bayeswave deglitching analysis, but we found this to be negligible compared to the parameter estimation itself.
Therefore, we can conclude that the Gaussian process analysis incurs significantly greater computational costs.
However, we remind the reader that this additional cost buys a more robust analysis, simultaneously analysing the signal and glitch.
In cases where \bayeswave finds difficulty in modelling the glitch (e.g. cases like the fast-scattering glitch where repeated artefacts are difficult to remove), it may be the only way to robustly determine the source properties.
To this end, we need to optimise the GP approach, for example, by developing custom jump proposals for the GP kernel terms, which will improve the efficiency of the \texttt{Bilby-MCMC} sampler and investigate explicit marginalisation.
Moreover, the discussion above centres on the stationary kernels available in \celerite.
When developing non-stationary kernels, it will be crucial to simultaneously find ways to ensure the per-likelihood evaluation time does not dramatically increase (though some extra cost is inevitable).

Third, we have not yet demonstrated that the GP inferences of the astrophysical signal parameters are unbiased, only that we can recover the true value of a simulated signal.
In future work, we will need to perform a parameter-parameter (PP) test \citep{cook06, talts18} a standard in the field to illustrate the posterior is unbiased (see, e.g., \cite{Veitch:2014wba}).
Such a test is complicated.
First, we to demonstrate the method is unbiased in stationary Gaussian noise, but more importantly, we then need to show it is unbiased for signals contaminated by a variety of glitches.
Choosing these glitches and how they contaminate the simulated signals in the PP test will require extensive investigation.

The three elements listed above constitute a minimum requirement to demonstrate that the Gaussian process approaches are a safe, efficient, and robust alternative to existing methods.
However, we expect there are many more improvements that could be made.
With the observational era of gravitational-wave astronomy firmly underway, now is an excellent time to explore new ideas in analysing the hundreds of events that are soon to be seen.
The Gaussian process approach fundamentally differs from existing approaches and offers new opportunities to think about how we analyse gravitational-wave events and test General Relativity.

\section{Data Availability}
The data used in the case studies of this work is available through the Gravitational Wave Open Science Center \citep{LIGOScientific:2019lzm}.

\section{Acknowledgements}

We would like to sincerely thank Dan Foreman-Mackay for help with the use of the \celerite Gaussian process python package, Colm Talbot for the suggestion of the Order Statistics approach to resolving the label-switching degeneracy, Walter Del Pozzo, Christopher Berry, and Andrew Lundgren for useful feedback on the manuscript, and Walter Del Pozzo, Marta Colleoni, Abhirup Ghosh, and Nathan Johnson-McDaniel for helpful feedback on the time-domain tests of GR.
This research has made use of data or software obtained from the Gravitational Wave Open Science Center (gw-openscience.org), a service of LIGO Laboratory, the LIGO Scientific Collaboration, the Virgo Collaboration, and KAGRA. LIGO Laboratory and Advanced LIGO are funded by the United States National Science Foundation (NSF) as well as the Science and Technology Facilities Council (STFC) of the United Kingdom, the Max-Planck-Society (MPS), and the State of Niedersachsen/Germany for support of the construction of Advanced LIGO and construction and operation of the GEO600 detector. Additional support for Advanced LIGO was provided by the Australian Research Council. Virgo is funded, through the European Gravitational Observatory (EGO), by the French Centre National de Recherche Scientifique (CNRS), the Italian Istituto Nazionale di Fisica Nucleare (INFN) and the Dutch Nikhef, with contributions by institutions from Belgium, Germany, Greece, Hungary, Ireland, Japan, Monaco, Poland, Portugal, Spain. The construction and operation of KAGRA are funded by Ministry of Education, Culture, Sports, Science and Technology (MEXT), and Japan Society for the Promotion of Science (JSPS), National Research Foundation (NRF) and Ministry of Science and ICT (MSIT) in Korea, Academia Sinica (AS) and the Ministry of Science and Technology (MoST) in Taiwan.

 This work makes use of the
\texttt{scipy} \citep{scipy_2020},
\texttt{numpy} \citep{harris_2020},
\texttt{matplotlib} \citep{Hunter:2007},
\texttt{gwpy} \citep{gwpy}, and
\texttt{PyCBC} \citep{nitz_2017} software for data analysis and visualisation.

\bibliographystyle{mnras}
\bibliography{bibliography}

\bsp
\label{lastpage}
\end{document}